\documentclass[prl,showpacs,twocolumn,floats,10pt,aps,citeautoscript,longbibliography,superscriptaddress]{revtex4-2}
\usepackage{dcolumn}
\usepackage{anyfontsize} % needed for \fontsize{}{}
\usepackage{microtype} % improves pdf generation
\usepackage{graphicx} 
\usepackage{xspace} % \xspace automatically adds space after string
\usepackage[usenames,dvipsnames]{xcolor} % text colors

\usepackage{xparse} % needed for NewDocumentCommand

\usepackage{algorithm}
\usepackage{algpseudocode}

\usepackage[normalem]{ulem} % to use strikethrough command \sout{}
\definecolor{darkgreen}{rgb}{0,0.5,0}
\definecolor{purple}{rgb}{0.6,0,0.5}
\definecolor{orange}{rgb}{1,0.5,0}
\definecolor{darkred}{rgb}{.7,0,0}
\definecolor{darkblue}{rgb}{0,0,.6}
\definecolor{grey}{rgb}{.6,.6,.6}
\definecolor{dimgreen}{rgb}{0.2,0.7,0.2}
\newcommand{\jvdomit}[1]{}
\usepackage{amsmath} 
\usepackage{amssymb}
\usepackage{mathrsfs} %needed for uppercase \ell: \eLL = \mathscr{L} 
\usepackage{bbm}  % double 1 for unit matrix \bbm{1}
\newcommand{\mr}[1]{\ensuremath{\mathrm{#1}}}
\newcommand{\mc}[1]{\ensuremath{\mathcal{#1}}}
\allowdisplaybreaks % page break in equations
\usepackage{enumitem} %to use enumerate environment with options

\usepackage{scalefnt} % needed for \scalefont{}

\usepackage{tikz}
\usepackage{algorithm}
\usepackage{algpseudocode}

\usepackage[colorlinks, citecolor={blue!50!black}, urlcolor={blue!50!black}, linkcolor={red!50!black}]{hyperref}
\usepackage{scalerel,stackengine}  % defining bracket \pig between \big and \Big

\newcommand{\ellplusone}{{\ell  \hspace{-0.2mm}+\hspace{-0.2mm} 1}}
\newcommand{\ellplustwo}{{\ell  \hspace{-0.2mm}+\hspace{-0.2mm} 2}}

\newcommand{\eLL}{{\mbox{\small$\mathscr{L}$}}}
\newcommand{\seLL}{{\scriptscriptstyle \! \mathscr{L}}}
\newcommand{\scripteLL}{{\scriptstyle \! \mathscr{L}}}

\newcommand{\mi}{\mathrm{i}} % imaginary unit
\newcommand*{\ndots}{\kern-0.075em.\kern-0.05em.\kern-0.05em.}  % narrow lower dots
\newcommand*{\nidots}{.\kern-0.05em.\kern-0.05em.} % narrow lower dots in subscript
\newcommand*{\ncdots}{\kern-0.15em\cdot\kern-0.2em\cdot\kern-0.2em\cdot\kern-0.15em}  % narrow centraldots
\newcommand{\bra}[1]{\ensuremath{\langle #1 |}}
\newcommand{\ket}[1]{\ensuremath{| #1 \rangle}}

\newcommand{\onesite}{\textrm{1s}}
\newcommand{\twosite}{\textrm{2s}}

\newcommand{\mbond}{\mathrm{b}}

\NewDocumentCommand{\doubleI}{O{}}{\mathbbm{1}_{#1}}
\NewDocumentCommand{\doubleIb}{O{}}{{\overline{\mathbbm{1}}_{#1}}}
\NewDocumentCommand{\doubleIk}{O{}}{\mathbbm{1}^\ks_{\! #1}}
\NewDocumentCommand{\doubleId}{O{}}{\mathbbm{1}^\ds_{\! #1}}
\NewDocumentCommand{\doubleIp}{O{}}{\mathbbm{1}^\ps_{\! #1}}
\NewDocumentCommand{\doubleV}{O{}}{\mathbbm{V}_{\! #1}}
\NewDocumentCommand{\doubleVk}{O{}}{\mathbbm{V}^\ks_{\! #1}}
\NewDocumentCommand{\doubleVd}{O{}}{\mathbbm{V}^\ds_{\! #1}}
\NewDocumentCommand{\doubleVp}{O{}}{\mathbbm{V}^\ps_{\! #1}}
\NewDocumentCommand{\doublev}{o}{{\mathbbm{v}_{#1}}}
\NewDocumentCommand{\doubleVb}{o}{{\overline{\mathbbm{V}}_{\! #1}}}
\NewDocumentCommand{\doubleVt}{o}{{\widetilde{\mathbbm{V}}_{\! #1}}}
\NewDocumentCommand{\doubleVh}{o}{\widehat{{\mathbbm{V}}_{\! #1}}}
\NewDocumentCommand{\doubleW}{o}{\mathbbm{W}_{\! #1}}
\NewDocumentCommand{\doubleWk}{o}{\mathbbm{W}^\ks_{\! #1}}
\NewDocumentCommand{\doubleWd}{o}{\mathbbm{W}^\ds_{\! #1}}
\NewDocumentCommand{\doubleWb}{o}{{\overline{\mathbbm{W}}_{\! #1}}}
\NewDocumentCommand{\doubleWt}{o}{{\widetilde{\mathbbm{V}}_{\! #1}}}
\NewDocumentCommand{\doubleWh}{o}{{\widehat{\mathbbm{V}}_{\! #1}}}

\def\D{{\scriptstyle {\rm D}}} % discarded
\def\DD{{\scriptstyle {\rm DD}}} % discarded discarded
\def\ds{{\scriptscriptstyle {\rm D}}}
\def\ks{{\scriptscriptstyle {\rm K}}}
\def\ps{{\scriptscriptstyle {\rm P}}}

\def\ps{{\scriptscriptstyle {\rm P}}}

\NewDocumentCommand{\cor}{mod()}% correlator
{
	#1\IfValueTF{#2}{[#2]}{}\IfValueTF{#3}{(#3)}{}
}

\usepackage{mathtools}
\usepackage{xparse} % needed to run NewDocumentCommand used in \xA, \xB, \xC

\NewDocumentCommand{\xA}{
         O{fill=black}mm D<|{0} D|>{0} D//{0} O{}O{0}O{0}}  
{
	\draw [#1] (#2-00.135,#3) -- (#2,#3) -- (#2,-00.135+#3) -- cycle;
	\draw (#2-00.25-#4,#3) -- (#2-00.135,#3); 
	\draw (#2,#3) -- (#2+00.25+#5,#3); 
	\draw (#2,#3-00.135) -- (#2,#3-00.25-#6);
	\draw (#2+#8,#3+#9) node (X) {#7};
}

\NewDocumentCommand{\xAd}{
         O{fill=black}mm D<|{0} D|>{0} D//{0} O{}O{0}O{0}}  
{
	\draw [#1] (#2-00.135,#3) -- (#2,#3) -- (#2,#3+00.135) -- cycle;
	\draw (#2-00.25-#4,#3) -- (#2-00.135,#3); 
	\draw (#2,#3) -- (#2+00.25+#5,#3); 
	\draw (#2,#3+00.135) -- (#2,#3+00.25+#6);
	\draw (#2+#8,#3+#9) node (X) {#7};
}

\NewDocumentCommand{\xB}{
         O{fill=black}mm D<|{0} D|>{0} D//{0} O{}O{0}O{0}} 
{
	\draw [#1] (#2,#3) -- (#2+0.135,#3) -- (#2,-0.135+#3) -- cycle;
	\draw (#2-00.25-#4,#3) -- (#2,#3); 
	\draw (#2+0.135,#3) -- (#2+00.25+#5,#3); 
	\draw (#2,#3-0.135) -- (#2,#3-00.25-#6);
	\draw (#2+#8,#3+#9) node (X) {#7};
}

\NewDocumentCommand{\xBd}{
         O{fill=black}mm D<|{0} D|>{0} D//{0} O{}O{0}O{0}}  
{
	\draw [#1] (#2,#3) -- (#2+0.135,#3) -- (#2,#3+0.135) -- cycle;
	\draw (#2-00.25-#4,#3) -- (#2,#3); 
	\draw (#2+0.135,#3) -- (#2+00.25+#5,#3); 
	\draw (#2,#3+0.135) -- (#2,#3+00.25+#6);
	\draw (#2+#8,#3+#9) node (X) {#7};
}

\NewDocumentCommand{\xC}{
         O{fill=black}mm D<|{0} D|>{0} D//{0} O{}O{0}O{0}}  
{
	\draw [#1] (#2,#3) circle (0.065);
	\draw (#2-00.25-#4,#3) -- (#2-0.065,#3); 
	\draw (#2+0.065,#3) -- (#2+00.25+#5,#3); 
	\draw (#2,#3-0.065) -- (#2,#3-00.25-#6);
	\draw (#2+#8,#3+#9) node (X) {#7};
}

\NewDocumentCommand{\xCd}{
         O{fill=black}mm D<|{0} D|>{0} D//{0} O{}O{0}O{0}}  
{
	\draw [#1] (#2,#3) circle (0.065);
	\draw (#2-00.25-#4,#3) -- (#2-0.065,#3); 
	\draw (#2+0.065,#3) -- (#2+00.25+#5,#3); 
	\draw (#2,#3+0.065) -- (#2,#3+00.25+#6);
	\draw (#2+#8,#3+#9) node (X) {#7};
}

\NewDocumentCommand{\xW}{
         O{fill=black}mm D<|{0} D|>{0} D//{0} O{}O{0}O{0}}  
{
	\draw [#1] (#2-0.065,#3-0.065) rectangle (#2+0.065,#3+0.065);
	\draw (#2-00.25-#4,#3) -- (#2-0.065,#3); 
	\draw (#2+0.065,#3) -- (#2+00.25+#5,#3); 
	\draw (#2,#3-0.065) -- (#2,#3-00.25-#6);
	\draw (#2,#3+0.065) -- (#2,#3+00.25+#6);
	\draw (#2+#8,#3+#9) node (X) {#7};
}

\NewDocumentCommand{\lcurl}{mmm  O{}O{0}O{0} D<>{0}}   
{
	\draw (#1,#3) edge[out=180,in=-90] (#1-0.1+#7,#2*0.5+#3*0.5);
	\draw (#1-0.1+#7,#2*0.5+#3*0.5) edge[out=90,in=180] (#1,#2);
	\draw (#1+#5,0.5*#3+0.5*2+#6) node (X) {#4};
}

\NewDocumentCommand{\rcurl}{mmm  O{}O{0}O{0} D<>{0}}   
{
	\draw (#1,#3) edge[out=0,in=-90] (#1+0.1+#7,#2*0.5+#3*0.5);
	\draw (#1+0.1+#7,#2*0.5+#3*0.5) edge[out=90,in=0] (#1,#2);
	\draw (#1+#5,0.5*#2+0.5*#3+#6) node (X) {#4};
}

\NewDocumentCommand{\effL}{O{fill=black} mmm  O{$\,$}O{0}O{0} D<>{0}}   
{
	\draw (#2,#4) edge[out=180,in=-90] (#2-00.2+#8,#3*0.5+#4*0.5-0.1);
	\draw (#2-00.2+#8,#3*0.5+#4*0.5+0.1) edge[out=90,in=180] (#2,#3);
	\draw [#1](#2-00.2+#8,#3*0.5+#4*0.5+0.1) -- 
	(#2-00.2+#8,#3*0.5+#4*0.5-0.1) -- (#2-00.1+#8,#3*0.5+#4*0.5) -- cycle;
	\draw (#2-00.2+#8,#3*0.5+#4*0.5) -- (#2,#3*0.5+#4*0.5);
	\draw (#2+#6,#4+#7) node (X) {#5};
}

\NewDocumentCommand{\effR}{O{fill=black} mmm  O{$\,$}O{0}O{0} D<>{0}}   
{
	\draw (#2,#4) edge[out=0,in=-90] (#2+00.2+#8,#3*0.5+#4*0.5-0.1);
	\draw (#2+00.2+#8,#3*0.5+#4*0.5+0.1) edge[out=90,in=0] (#2,#3);
	\draw [#1] (#2+00.2+#8,#3*0.5+#4*0.5+0.1) -- 
	(#2+00.2+#8,#3*0.5+#4*0.5-0.1) -- (#2+00.1+#8,#3*0.5+#4*0.5) -- cycle;
	\draw (#2+00.2+#8,#3*0.5+#4*0.5) -- (#2,#3*0.5+#4*0.5);
	\draw (#2+#6,#4+#7) node (X) {#5};
}

\makeatletter
\def\maketitle{
\@author@finish
\title@column\titleblock@produce
\suppressfloats[t]}
\makeatother

\begin{document} 

\title{Reply to comment on ``Controlled bond expansion for Density Matrix Renormalization Group ground state search at single-site costs''
}
\author{Andreas Gleis}
\affiliation{Arnold Sommerfeld Center for Theoretical Physics, 
Center for NanoScience,\looseness=-1\,  and 
Munich Center for \\ Quantum Science and Technology,\looseness=-2\, 
Ludwig-Maximilians-Universit\"at M\"unchen, 80333 Munich, Germany}
\author{Jheng-Wei Li}
\affiliation{Arnold Sommerfeld Center for Theoretical Physics, 
Center for NanoScience,\looseness=-1\,  and 
Munich Center for \\ Quantum Science and Technology,\looseness=-2\, 
Ludwig-Maximilians-Universit\"at M\"unchen, 80333 Munich, Germany}
\author{Jan von Delft}
\affiliation{Arnold Sommerfeld Center for Theoretical Physics, 
Center for NanoScience,\looseness=-1\,  and 
Munich Center for \\ Quantum Science and Technology,\looseness=-2\, 
Ludwig-Maximilians-Universit\"at M\"unchen, 80333 Munich, Germany}

\begin{abstract}
\begin{center}
(Dated: \today)
\end{center}

\end{abstract}

\maketitle

In a recent comment~\cite{McCulloch2024}, McCulloch and Osborne~(MO)
offer constructive criticism of Ref.~\onlinecite{Gleis2023a}, where we introduced a  controlled bond expansion~(CBE) algorithm for DMRG ground state searches to achieve 2-site (2s) updates at single-site cost. They also discuss its convergence in comparison to the strictly single-site (3S) algorithm~\cite{Hubig2015} for expanding bonds. 
We agree with some but not all of their remarks, and advocate combining the CBE and 3S strategies to achieve very favorable convergence properties.

\textit{Variational properties.---} Both CBE-DMRG and 2s-DMRG evoke a variational principle---minimization of the energy expectation value---to derive a prescription for updating MPS tensors. In this loose sense, they are variational. Strictly 
speaking, however, they are not, since updated tensors are subsequently truncated, which may slightly increase the energy. Thus, steps (i-iii) of the CBE algorithm are variational but step (iv) is not. We thank MO for emphasizing this point, which 
we inadvertently disregarded several times in the phrasing used in Ref.~\cite{Gleis2023a}.

\textit{2s tangent space.---}
In their comment, MO state that the $\DD$ projection during CBE expansion is not useful or may even negatively affect the results, 
especially when used in the context of TDVP. We emphatically disagree.

From a general perspective, both ground state searches via 1-site (1s) DMRG  and time evolution via 1s-TDVP can be viewed as algorithms which modify a reference state $|\Psi\rangle$ 
by incorporating contributions from its 1s tangent space $\doubleV^{\onesite}$ (we use the notation of Ref.~\cite{Gleis2022a}). 
2-site (2s) DMRG and 2s-TDVP additionally incorporate information from the 2s tangent space $\doubleV^{
\twosite}$. CBE expands bonds by targeting states from $\doubleV^{2\perp} =  \doubleV^{\twosite} \backslash \doubleV^{\onesite}$ 
that contribute the most weight to $H \ket{\Psi}$.
Since $\doubleV^{2\perp}$ is the image of the projector  $\mc{P}^{2\perp} = \sum_{\ell =1}^{\seLL-1} \mc{P}_{\ell,\ell+1}^{\ds\ds}$ (see Eq.~(61) of Ref.~\cite{Gleis2022a}), this can be done \textit{unambiguously} by sequentially considering $\mc{P}_{\ell,\ell+1}^{\ds\ds} H \ket{\Psi}$---a choice dictated by the structure of the 1s and 2s tangent space projectors. 

\textit{Ad hoc} modifications of this choice, disregarding
said structure, may lead to inferior expansion vectors. MO's suggestion to omit the full $\DD$ projection $\mc{P}_{\ell,\ell+1}^{\ds\ds}$ is a case in point. In doing so, one would potentially 
target directions in $\doubleV^{\onesite}$ for bond expansion, which are already considered during single-site updates.
For DMRG, that would be computationally inefficient, but not compromise accuracy; for TDVP, however, it can lead to avoidable errors.
To solidify our statement, we provide concrete examples in the supplemental material~\cite{supp_MOcomment}.
\nocite{Stoudenmire2012,Haegeman2014,Haegeman2016,Hubig2018,Li2024a}
Since we are not aware of situations where the $\DD$ projection is problematic, we invite 
MO to solidify their claim with a concrete example.
The $\mc{O}(D^3 d)$ cost of the projection is negligible.

\textit{Randomized SVD.}--- MO's suggestion to replace ``shrewd selection" with randomized SVD in CBE is interesting!
It could be used to simplify the CBE algorithm by directly minimizing the cost function $\mathcal{C}_1$ 
of Fig.~1 in Ref.~\onlinecite{Gleis2023a}, without recourse
to $\mathcal{C}_2$ and $\mathcal{C}_3$.
We will try this ourselves and look forward to MO's results.

\textit{Comparison to 3S.}--- In Sec.~S-3 of Ref.~\cite{Gleis2023a}, we compared the convergence properties of CBE--DMRG to the strictly single-site (3S) algorithm~\cite{Hubig2015} for expanding bonds, reporting faster convergence for the former for the models studied there.  MO point out that 3S can converge in situations where the CBE-DMRG and 2s-DMRG algorithms do not. We agree. 
This reflects the fact that 3S and CBE/2s updates address two different issues arising in DMRG. The 3S mixing step~\cite{Hubig2015} aims 
to drive an MPS away from a local variational minimum to avoid getting stuck in some metastable, suboptimal solution. To this end, 3S perturbs the current MPS to add subspaces likely to reach other, deeper local minima in future sweeps~\cite{McCulloch2024}. In doing so, 3S may truncate subspaces useful within the current local minimum, causing a momentary increase in energy.  
By contrast, CBE/2s algorithms aim to add subspaces most relevant to lower the energy within the current local minimum during 
the current sweep. 
Thus, CBE/2s converge quickly to close-by (potentially local) minima. They also provide an intrinsic error measure to estimate the distance from the minimum.

Due to the complementary properties of 2s and mixing algorithms~\cite{White2005} 
it is fruitful to use both in combination~\cite{Stoudenmire2012}. In the same spirit, we view the 
CBE and 3S algorithms not as competitors but as allies,
and recommend using a \textit{combination} of CBE, yielding 2s updates at 1s costs, 
 and strong 3S mixing. We called this scheme CBE$+\alpha$ in Sec.~S-3 of Ref.~\onlinecite{Gleis2023a}. On the one hand, 3S with a very large(!) mixing parameter of $\alpha = \mc{O}(1)$ facilitates efficiently 
escaping from metastable minima.
On the other hand, CBE yields a rapid descent
  to the bottom of a close-by accessible minimum. 
   Thus, their combination CBE+$\alpha$ yields very favorable convergence properties, illustrated in the supplement~\cite{supp_MOcomment} of this Reply. 
  There, we also reply to MO's remarks on Figs.~S-7 and S-9 of Ref.~\onlinecite{Gleis2023a}, comparing the convergence of CBE and 3S.

\bibliography{CBE-DMRG}

\begin{thebibliography}{11}%
\makeatletter
\providecommand \@ifxundefined [1]{%
 \@ifx{#1\undefined}
}%
\providecommand \@ifnum [1]{%
 \ifnum #1\expandafter \@firstoftwo
 \else \expandafter \@secondoftwo
 \fi
}%
\providecommand \@ifx [1]{%
 \ifx #1\expandafter \@firstoftwo
 \else \expandafter \@secondoftwo
 \fi
}%
\providecommand \natexlab [1]{#1}%
\providecommand \enquote  [1]{``#1''}%
\providecommand \bibnamefont  [1]{#1}%
\providecommand \bibfnamefont [1]{#1}%
\providecommand \citenamefont [1]{#1}%
\providecommand \href@noop [0]{\@secondoftwo}%
\providecommand \href [0]{\begingroup \@sanitize@url \@href}%
\providecommand \@href[1]{\@@startlink{#1}\@@href}%
\providecommand \@@href[1]{\endgroup#1\@@endlink}%
\providecommand \@sanitize@url [0]{\catcode `\\12\catcode `\$12\catcode
  `\&12\catcode `\#12\catcode `\^12\catcode `\_12\catcode `\%12\relax}%
\providecommand \@@startlink[1]{}%
\providecommand \@@endlink[0]{}%
\providecommand \url  [0]{\begingroup\@sanitize@url \@url }%
\providecommand \@url [1]{\endgroup\@href {#1}{\urlprefix }}%
\providecommand \urlprefix  [0]{URL }%
\providecommand \Eprint [0]{\href }%
\providecommand \doibase [0]{https://doi.org/}%
\providecommand \selectlanguage [0]{\@gobble}%
\providecommand \bibinfo  [0]{\@secondoftwo}%
\providecommand \bibfield  [0]{\@secondoftwo}%
\providecommand \translation [1]{[#1]}%
\providecommand \BibitemOpen [0]{}%
\providecommand \bibitemStop [0]{}%
\providecommand \bibitemNoStop [0]{.\EOS\space}%
\providecommand \EOS [0]{\spacefactor3000\relax}%
\providecommand \BibitemShut  [1]{\csname bibitem#1\endcsname}%
\let\auto@bib@innerbib\@empty
%</preamble>
\bibitem [{\citenamefont {McCulloch}\ and\ \citenamefont
  {Osborne}(2024)}]{McCulloch2024}%
  \BibitemOpen
  \bibfield  {author} {\bibinfo {author} {\bibfnamefont {I.~P.}\ \bibnamefont
  {McCulloch}}\ and\ \bibinfo {author} {\bibfnamefont {J.}~\bibnamefont
  {Osborne}},\ }\bibfield  {title} {\bibinfo {title} {Comment on "controlled
  bond expansion for density matrix renormalization group ground state search
  at single-site costs"},\ }\href@noop {} {\  (\bibinfo {year}
  {2024})}\BibitemShut {NoStop}%
\bibitem [{\citenamefont {Gleis}\ \emph {et~al.}(2023)\citenamefont {Gleis},
  \citenamefont {Li},\ and\ \citenamefont {von Delft}}]{Gleis2023a}%
  \BibitemOpen
  \bibfield  {author} {\bibinfo {author} {\bibfnamefont {A.}~\bibnamefont
  {Gleis}}, \bibinfo {author} {\bibfnamefont {J.-W.}\ \bibnamefont {Li}},\ and\
  \bibinfo {author} {\bibfnamefont {J.}~\bibnamefont {von Delft}},\ }\bibfield
  {title} {\bibinfo {title} {Controlled bond expansion for density matrix
  renormalization group ground state search at single-site costs},\ }\href
  {https://doi.org/10.1103/PhysRevLett.130.246402} {\bibfield  {journal}
  {\bibinfo  {journal} {Phys. Rev. Lett.}\ }\textbf {\bibinfo {volume} {130}},\
  \bibinfo {pages} {246402} (\bibinfo {year} {2023})}\BibitemShut {NoStop}%
\bibitem [{\citenamefont {Hubig}\ \emph {et~al.}(2015)\citenamefont {Hubig},
  \citenamefont {McCulloch}, \citenamefont {Schollw\"ock},\ and\ \citenamefont
  {Wolf}}]{Hubig2015}%
  \BibitemOpen
  \bibfield  {author} {\bibinfo {author} {\bibfnamefont {C.}~\bibnamefont
  {Hubig}}, \bibinfo {author} {\bibfnamefont {I.~P.}\ \bibnamefont
  {McCulloch}}, \bibinfo {author} {\bibfnamefont {U.}~\bibnamefont
  {Schollw\"ock}},\ and\ \bibinfo {author} {\bibfnamefont {F.~A.}\ \bibnamefont
  {Wolf}},\ }\bibfield  {title} {\bibinfo {title} {Strictly single-site {DMRG}
  algorithm with subspace expansion},\ }\href
  {https://doi.org/10.1103/PhysRevB.91.155115} {\bibfield  {journal} {\bibinfo
  {journal} {Phys. Rev. B}\ }\textbf {\bibinfo {volume} {91}},\ \bibinfo
  {pages} {155115} (\bibinfo {year} {2015})}\BibitemShut {NoStop}%
\bibitem [{\citenamefont {Gleis}\ \emph {et~al.}(2022)\citenamefont {Gleis},
  \citenamefont {Li},\ and\ \citenamefont {von Delft}}]{Gleis2022a}%
  \BibitemOpen
  \bibfield  {author} {\bibinfo {author} {\bibfnamefont {A.}~\bibnamefont
  {Gleis}}, \bibinfo {author} {\bibfnamefont {J.-W.}\ \bibnamefont {Li}},\ and\
  \bibinfo {author} {\bibfnamefont {J.}~\bibnamefont {von Delft}},\ }\bibfield
  {title} {\bibinfo {title} {Projector formalism for kept and discarded spaces
  of matrix product states},\ }\href
  {https://doi.org/10.1103/PhysRevB.106.195138} {\bibfield  {journal} {\bibinfo
   {journal} {Phys. Rev. B}\ }\textbf {\bibinfo {volume} {106}},\ \bibinfo
  {pages} {195138} (\bibinfo {year} {2022})}\BibitemShut {NoStop}%
\bibitem [{sup()}]{supp_MOcomment}%
  \BibitemOpen
  \href@noop {} {\bibinfo  {journal} {See Supplemental Material at [url] for
  more details on CBE$+\alpha$, an additional comparison of CBE, 3S, and
  CBE$+\alpha$, and a discussion on the importance of projecting to the 2-site
  tangent space in CBE. The Supplemental Material includes
  Refs.~\cite{Gleis2023a,Stoudenmire2012,McCulloch2024,Hubig2015,Gleis2022a,Li2024a,Haegeman2014,Haegeman2016,Hubig2018}}\
  }\BibitemShut {NoStop}%
\bibitem [{\citenamefont {Stoudenmire}\ and\ \citenamefont
  {White}(2012)}]{Stoudenmire2012}%
  \BibitemOpen
\bibfield  {journal} {  }\bibfield  {author} {\bibinfo {author} {\bibfnamefont
  {E.}~\bibnamefont {Stoudenmire}}\ and\ \bibinfo {author} {\bibfnamefont
  {S.~R.}\ \bibnamefont {White}},\ }\bibfield  {title} {\bibinfo {title}
  {Studying two-dimensional systems with the density matrix renormalization
  group},\ }\href {https://doi.org/10.1146/annurev-conmatphys-020911-125018}
  {\bibfield  {journal} {\bibinfo  {journal} {Ann. Rev. Cond. Mat. Phys.}\
  }\textbf {\bibinfo {volume} {3}},\ \bibinfo {pages} {111} (\bibinfo {year}
  {2012})}\BibitemShut {NoStop}%
\bibitem [{\citenamefont {Haegeman}\ \emph {et~al.}(2014)\citenamefont
  {Haegeman}, \citenamefont {Mariën}, \citenamefont {Osborne},\ and\
  \citenamefont {Verstraete}}]{Haegeman2014}%
  \BibitemOpen
  \bibfield  {author} {\bibinfo {author} {\bibfnamefont {J.}~\bibnamefont
  {Haegeman}}, \bibinfo {author} {\bibfnamefont {M.}~\bibnamefont {Mariën}},
  \bibinfo {author} {\bibfnamefont {T.~J.}\ \bibnamefont {Osborne}},\ and\
  \bibinfo {author} {\bibfnamefont {F.}~\bibnamefont {Verstraete}},\ }\bibfield
   {title} {\bibinfo {title} {{Geometry of matrix product states: Metric,
  parallel transport, and curvature}},\ }\href
  {https://doi.org/10.1063/1.4862851} {\bibfield  {journal} {\bibinfo
  {journal} {Journal of Mathematical Physics}\ }\textbf {\bibinfo {volume}
  {55}},\ \bibinfo {pages} {021902} (\bibinfo {year} {2014})}\BibitemShut
  {NoStop}%
\bibitem [{\citenamefont {Haegeman}\ \emph {et~al.}(2016)\citenamefont
  {Haegeman}, \citenamefont {Lubich}, \citenamefont {Oseledets}, \citenamefont
  {Vandereycken},\ and\ \citenamefont {Verstraete}}]{Haegeman2016}%
  \BibitemOpen
  \bibfield  {author} {\bibinfo {author} {\bibfnamefont {J.}~\bibnamefont
  {Haegeman}}, \bibinfo {author} {\bibfnamefont {C.}~\bibnamefont {Lubich}},
  \bibinfo {author} {\bibfnamefont {I.}~\bibnamefont {Oseledets}}, \bibinfo
  {author} {\bibfnamefont {B.}~\bibnamefont {Vandereycken}},\ and\ \bibinfo
  {author} {\bibfnamefont {F.}~\bibnamefont {Verstraete}},\ }\bibfield  {title}
  {\bibinfo {title} {Unifying time evolution and optimization with matrix
  product states},\ }\href {https://doi.org/10.1103/PhysRevB.94.165116}
  {\bibfield  {journal} {\bibinfo  {journal} {Phys. Rev. B}\ }\textbf {\bibinfo
  {volume} {94}},\ \bibinfo {pages} {165116} (\bibinfo {year}
  {2016})}\BibitemShut {NoStop}%
\bibitem [{\citenamefont {Hubig}\ \emph {et~al.}(2018)\citenamefont {Hubig},
  \citenamefont {Haegeman},\ and\ \citenamefont {Schollw\"ock}}]{Hubig2018}%
  \BibitemOpen
  \bibfield  {author} {\bibinfo {author} {\bibfnamefont {C.}~\bibnamefont
  {Hubig}}, \bibinfo {author} {\bibfnamefont {J.}~\bibnamefont {Haegeman}},\
  and\ \bibinfo {author} {\bibfnamefont {U.}~\bibnamefont {Schollw\"ock}},\
  }\bibfield  {title} {\bibinfo {title} {Error estimates for extrapolations
  with matrix-product states},\ }\href
  {https://doi.org/10.1103/PhysRevB.97.045125} {\bibfield  {journal} {\bibinfo
  {journal} {Phys. Rev. B}\ }\textbf {\bibinfo {volume} {97}},\ \bibinfo
  {pages} {045125} (\bibinfo {year} {2018})}\BibitemShut {NoStop}%
\bibitem [{\citenamefont {Li}\ \emph {et~al.}(2024)\citenamefont {Li},
  \citenamefont {Gleis},\ and\ \citenamefont {von Delft}}]{Li2024a}%
  \BibitemOpen
  \bibfield  {author} {\bibinfo {author} {\bibfnamefont {J.-W.}\ \bibnamefont
  {Li}}, \bibinfo {author} {\bibfnamefont {A.}~\bibnamefont {Gleis}},\ and\
  \bibinfo {author} {\bibfnamefont {J.}~\bibnamefont {von Delft}},\ }\bibfield
  {title} {\bibinfo {title} {Time-dependent variational principle with
  controlled bond expansion for matrix product states},\ }\href
  {https://doi.org/10.1103/PhysRevLett.133.026401} {\bibfield  {journal}
  {\bibinfo  {journal} {Phys. Rev. Lett.}\ }\textbf {\bibinfo {volume} {133}},\
  \bibinfo {pages} {026401} (\bibinfo {year} {2024})}\BibitemShut {NoStop}%
\bibitem [{\citenamefont {White}(2005)}]{White2005}%
  \BibitemOpen
  \bibfield  {author} {\bibinfo {author} {\bibfnamefont {S.~R.}\ \bibnamefont
  {White}},\ }\bibfield  {title} {\bibinfo {title} {Density matrix
  renormalization group algorithms with a single center site},\ }\href
  {https://doi.org/10.1103/PhysRevB.72.180403} {\bibfield  {journal} {\bibinfo
  {journal} {Phys. Rev. B}\ }\textbf {\bibinfo {volume} {72}},\ \bibinfo
  {pages} {180403} (\bibinfo {year} {2005})}\BibitemShut {NoStop}%
\end{thebibliography}%

\clearpage

\newpage

\title{Supplemental material: \\ Reply to comment on ``Controlled bond expansion for Density Matrix Renormalization Group ground state search at single-site costs''}

\date{\today}
\maketitle

\setcounter{secnumdepth}{2} % enable section numbering, which is disabled in prl
\renewcommand{\thefigure}{S-\arabic{figure}}% change figure numbering style for appendix
\setcounter{figure}{0}
\setcounter{section}{0}
\setcounter{equation}{0}
\renewcommand{\thesection}{S-\arabic{section}}% change section numbering style for appendix
\renewcommand{\theequation}{S\arabic{equation}}% change equation numbering style for appendix

In Section~\ref{sec:CBE_3S_compare}, we 
give a step-by-step definition of the CBE$+\alpha$ algorithm,
explain how we use it while sweeping, and
provide an additional comparison of CBE, 3S, and CBE$+\alpha$ 
(complementing that given in Sec.~S-3 of Ref.~\onlinecite{Gleis2023a})
to illustrate the convergence properties of these methods.In Section~\ref{sec:2siteProjection}, we explain in more detail why we perform the $\DD$ projection in CBE
and how this affects the results, especially in TDVP. 

\vspace{-0.5cm}
\section{Comparison of CBE, 3S and CBE$+\alpha$}
\label{sec:CBE_3S_compare}
\vspace{-0.2cm}

As we have stated in the main text, mixing techniques like 3S and CBE/2s aim at solving different issues of DMRG. 
Thus, it is in our opinion best practice to combine those techniques, c.f., for instance, Ref.~\onlinecite{Stoudenmire2012} or our recommended
CBE$+\alpha$ in Sec.~S-4 of Ref.~\onlinecite{Gleis2023a}.
On the one hand, this prevents CBE/2s from getting stuck in local minima. On the other hand, it avoids the cumbersome task of converging an MPS using a mixing technique
like 3S by fine-tuning the mixing parameter $\alpha$. Further, it allows the use of a large $\alpha = \mc{O}(1)$, which is in our view required to get a calculation stuck in a local minimum going again by escaping a metastable solution. 

The CBE+$\alpha$ algorithm is defined as follows. 
Consider the update of site $\ell+1$ during a right-to-left sweep. As also pointed out in MO's arXiv comment~\cite{McCulloch2024}, 
CBE expands the bond between sites $\ell$ and $\ell+1$ \textit{before} updating site $\ell+1$ (i.e.\ CBE does pre-expansion~\cite{McCulloch2024}), while 3S mixing on the same bond
 is performed \textit{after} updating site $\ell+1$ (i.e.\ 3S does post-expansion~\cite{McCulloch2024}).
 In a right-to-left CBE$+\alpha$ sweep, we proceed as follows to update site $\ell+1$:
 (i) Compute the truncated complement $\widetilde{A}^{\mr{tr}}_{\ell}$.  
 (ii) Use it to obtain the expanded isometry $A^{\mr{ex}}_{\ell}$, the expanded isometry center $C^{\mr{ex}}_{\ell+1}$ 
 and the expanded one-site Hamiltonian $H^{\onesite,\mr{ex}}_{\ell+1}$ [steps (i) and (ii) in Ref.~\onlinecite{Gleis2023a}].
 (iii) Update $C^{\mr{ex}}_{\ell+1}$ by (approximately) finding the ground-state of $H^{\onesite,\mr{ex}}_{\ell+1}$ [step (iii) in Ref.~\onlinecite{Gleis2023a}].
 (iv) Truncate $C^{\mr{ex}}_{\ell+1}$ using SVD, but do not shift the isometry center.
 (v) Perform 3S mixing with mixing parameter $\alpha$ on the bond between sites $\ell+1$ and $\ell$, according to Eq.~(S4) of Ref.~\onlinecite{Gleis2023a}
 (which transcribes the 3S algorithm of Ref.~\onlinecite{Hubig2015} into the notation used in Ref.~\onlinecite{Gleis2023a}).
 After step (v), the isometry center is located at site $\ell$. If $\alpha = 0$, step (v) is omitted and the isometry center is shifted in step (iv). 

In our applications, we typically use CBE$+\alpha$ with $\alpha = \mc{O}(1)$ during the bond-growing phase
and then set $\alpha = 0$ for a few sweeps before obtaining error measures and measuring observables.
When we perform a series of computations at different bond dimensions $D_1 < D_2,\dots$ to monitor convergence, we typically first converge an MPS at $D_1$.
After obtaining error measures and measuring observables, we converge a $D_2$ MPS using the previous $D_1$ MPS as initialization and performing 
typically one or two sweeps at $\alpha = \mc{O}(1)$, followed by a few sweeps with $\alpha = 0$. Larger $D_n$ are reached correspondingly. 
We find that our strategy leads to quick convergence while avoiding local minima.

% Figure1
\begin{figure}
\includegraphics[width = \linewidth]{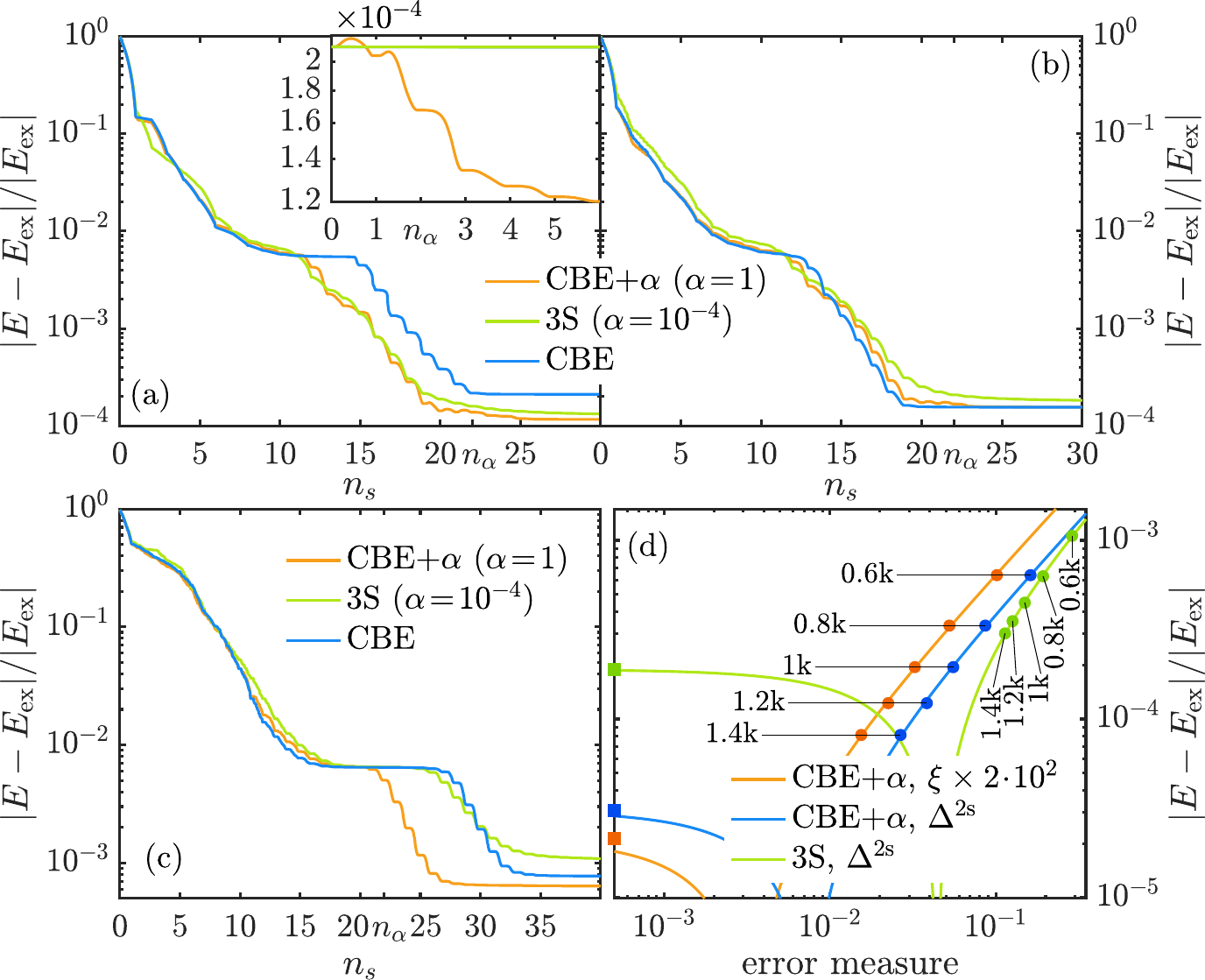}
\caption{\label{fig:FFPBC_FFCyl} 
(a-d) Convergence of the GS energy versus number of half-sweeps $n_s$ using different bond expansion methods at $D^{\ast}_{\mr{max}} = 600$. 
Spinful free fermions on a $\eLL = 100$ ring with nearest-neighbor hopping at (a) half-filling, (b) filling $N = 0.9\eLL$. 
Inset in (a): attempt to converge the CBE MPS using pure 3S and CBE$+\alpha$ (see text for details). 
(c) Spinful free fermions on a $\eLL_x \times \eLL_y = 10\times 4$ cylinder
with nearest-neighbor hopping at filling $N=0.9\eLL_x \eLL_y$. 
For CBE$+\alpha$, we used $\alpha = 1$ for $n_s \leq n_{\alpha} = 22$ and $\alpha=0$ for $n_s > n_{\alpha}$. 
(d) Energy versus error measure ($\Delta^{2\mr{s}}$: 2s variance, $\xi$: discarded weight) for CBE$+\alpha$ and 3S at different $D^{\ast}_{\mr{max}}$.
Dots are data points, lines linear fits and squares indicate the extrapolated energy. 
\vspace{-0.6cm}
}
\end{figure}

\textit{Results.---}
In the following, we illustrate our discussion above based on different examples. All of our examples are spinful fermionic models and we exploit SU(2) spin rotation $\times$ U(1) charge symmetry.
We compare pure CBE, CBE$+\alpha$ with $\alpha = 1$ and pure 3S with $\alpha = 10^{-4}$. In CBE$+\alpha$, we set $\alpha=1$ up to and including sweep number $n_{\alpha}$ and $\alpha=0$ afterwards.
If not stated otherwise, the maximum bond dimension is $D^{\ast}_{\mr{max}} = 600$, we start with a $D^{\ast} = 1$ initial state and we grow $D^{\ast}$ by $\sqrt{2}$ every half-sweep until $D^{\ast}_{\mr{max}}$
is reached.

Our first example considers half-filled spinful free fermions on a ring with nearest-neighbor hopping, the same model as used in Ref.~\onlinecite{McCulloch2024}, Fig.~1.
Figure~\ref{fig:FFPBC_FFCyl}~(a) shows our results regarding convergence of different methods.
As pointed out in Ref.~\onlinecite{McCulloch2024}, pure CBE (blue, not recommended!) first converges towards 
a metastable minimum (corresponding to the open-boundary solution), then lowers
the energy further and finally gets stuck at an error which is almost twice as large as the one achievable with CBE$+\alpha$ (orange). 
Both CBE$+\alpha$ and 3S (green) converge nicely, with CBE$+\alpha$ converging slightly better.

To test the capability of the different methods to escape local minima, we initialize them with the $D^{\ast} = 600$ MPS obtained from pure CBE.
We then perform 6 half-sweeps with both 3S (using $\alpha = 10^{-4}$) and CBE$+\alpha$; for the latter, we use $\alpha = 1$ for the first two half-sweeps and then set $\alpha=0$.
We also check pure CBE for reference. The results are shown in the inset of Fig.~\ref{fig:FFPBC_FFCyl}~(a). Clearly, only CBE$+\alpha$ 
succeeds in escaping the local minimum, while 3S and CBE remain stuck (their curves coincide). This clearly illustrates the issues that can 
arise when using small $\alpha = 10^{-4}$, as required in pure 3S calculations.

In Fig.~\ref{fig:FFPBC_FFCyl}(b), we show the same free fermion ring, but now filled with $N = 0.9\eLL = 90$ Fermions. Pure CBE converges now slightly better
and to a slightly lower energy than 3S. This illustrates that long-range terms in the Hamiltonian are not generically an issue for pure CBE. 
CBE$+\alpha$ converges without issues.

Figure~\ref{fig:FFPBC_FFCyl}(c) shows convergence versus sweeps for spinful free fermions with nearest-neighbor hopping on a $\eLL_x \times \eLL_y = 10\times4$ cylinder
at filling $N = 0.9\eLL_x \eLL_y = 36$. Clearly, CBE$+\alpha$ converges in the fewest number of sweeps and to the lowest energy. 3S and CBE both get stuck for some 
time in a local minimum, which both methods manage to escape in a similar fashion. Pure CBE then converges faster than 3S and to a lower energy. 
We emphasize that pure 3S has similar convergence problems for this model as pure CBE had for the model in Fig.~\ref{fig:FFPBC_FFCyl}~(a),
converging to an error almost twice as large as pure CBE or CBE$+\alpha$. This illustrates that also pure 3S can have serious convergence problems
and further emphasizes the stable performance of CBE$+\alpha$.

In Fig.~\ref{fig:FFPBC_FFCyl}(d), we illustrate the importance of quick convergence in a setting where we intend to converge a ground state of a model and monitor the convergence. 
For that, we compute a series of ground states with different $D^{\ast}_n \in \{600,800,1000,1200,1400\}$ for the free-fermion cylinder from the previous paragraph,
using both 3S and CBE$+\alpha$. We first converge the MPS with the corresponding method at $D^{\ast}_1 = 600$ as shown in Fig.~\ref{fig:FFPBC_FFCyl}~(c)
and then successively increase the bond dimension. In between, we perform 6 half-sweeps with either 3S or CBE$+\alpha$. For the latter, we set $\alpha=1$
during the first two half-sweeps and then use $\alpha=0$. At every $D^{\ast}_n$, we compute the 2s variance~\cite{Hubig2018} $\Delta^{2\mr{s}}$ as an error measure for both 3S and CBE$+\alpha$;
for the latter, the discarded weight $\xi$ from CBE is also a readily available error measure, and the computation of $\Delta^{2\mr{s}}$ would usually be omitted. 

Figure~\ref{fig:FFPBC_FFCyl}(d) shows the energies at different $D^{\ast}_n$ versus error measure (dots) and a corresponding linear extrapolation (lines) to zero error measure (squares) to estimate the error.
It is evident that CBE$+\alpha$ performs far better than 3S. At $D^{\ast} = 1400$, 3S manages to barely surpass the accuracy obtained by CBE$+\alpha$ already at $D^{\ast} = 800$, 
i.e.\ at almost half of the bond dimension for the latter. 
The corresponding CBE$+\alpha$ result at $D^{\ast} = 1400$ has an error that is almost 4 times lower than the corresponding 3S result. After extrapolation, 3S reaches an error which is almost identical 
to that of the $D^{\ast} = 1000$ CBE$+\alpha$ calculation, the $D^{\ast} \in \{1200,1400\}$ CBE$+\alpha$ calculations clearly surpass the extrapolated 3S result in accuracy and the extrapolated CBE$+\alpha$ 
results are an order of magnitude more accurate than the corresponding 3S result. 

To conclude this comparison, we have illustrated that CBE$+\alpha$ (recommended in our Letter~\cite{Gleis2023a}) is an algorithm with exceptional convergence properties.
Its ability to escape or avoid local minima clearly surpasses that of pure 3S because a significantly larger mixing parameter $\alpha = \mc{O}(1)$ can be used. 
On the other hand, it inherits the favorable convergence properties of CBE close to (local) minima. 
This becomes particularly clear in Fig.~\ref{fig:FFPBC_FFCyl}~(d), which also shows that the convergence issues of 3S cannot be simply cured 
by somewhat increasing $D^{\ast}$, in contrast to what MO claim in Ref.~\onlinecite{McCulloch2024}: when increasing $D^{\ast}$,
3S falls even further behind CBE$+\alpha$ in terms of accuracy. 

For completeness, we now address the specific remarks of MO at the end of their comment.
MO further suggest that for a fair comparison, $D(1+\delta)$ states should be kept in 3S, i.e. the same number of states considered during the CBE optimization before truncation. 
We are not entirely sure what MO mean by that. 3S expands by $Dw \gg D \delta$ states, i.e. by much more than CBE. Both CBE and 3S truncate back to bond dimension $D$,
such that the final state is a bond dimension $D$ MPS, so the comparison seems fair to us. The main difference is that 3S by construction does not optimize the wavefunction
directly after expansion, but first truncates before optimization. 
We are not sure whether MO suggest switching the order of truncation and optimization in 3S. 

We note that when comparing a bond-dimension $D(1+0.1)$ 3S calculation with a bond-dimension $D$ CBE$+\alpha$ calculation, 3S usually does not 
provide a better energy, see for instance Fig.~\ref{fig:FFPBC_FFCyl}(d) and our discussion above. 
We also do not see how this follows from our Fig.~S-7 of Ref.~\onlinecite{Gleis2023a}.

It is indeed true that 2s DMRG can also converge for the calculation in Fig.~S-9(a) of Ref.~\cite{Gleis2023a} if singular vectors with zero singular values are kept.
However, using CBE$+\alpha$ yields much quicker convergence.

\vspace{-0.5cm}
\section{2s tangent space}
\label{sec:2siteProjection}
\vspace{-0.2cm}

In their comment, MO claim that the $\DD$ projection we perform to select candidate vectors for bond expansion is 
not useful and can even lead to errors, especially in CBE--TDVP~\cite{Li2024a}. 
We strongly disagree with this statement.

To explain why, we first provide an overview of the role of the 2s tangent space in DMRG and TDVP in Sec.~\ref{subsec:RoleOf2s}.
This section is intended to provide a brief summary of (our view on) the general philosophy behind DMRG and TDVP 
and how the 2s or CBE update fits into this philosophy. 
In Sec.~\ref{subsec:DDimportance}, we explain in more detail why the $\DD$ projection is necessary for a successful CBE update.
There, we also provide an explicit example from CBE--TDVP which illustrates what can go wrong when omitting the full $\DD$ projection.
Below, we extensively use concepts and notation introduced in Ref.~\onlinecite{Gleis2022a}. 

\vspace{-0.5cm}
\subsection{Role of 2s tangent space --- DMRG and TDVP}
\label{subsec:RoleOf2s}
\vspace{-0.2cm}

First, consider the single-site DMRG or TDVP update for a given non-uniform MPS $\ket{\Psi}$.
During the single-site update, changes of $\ket{\Psi}$ are considered in the tangent space $\doubleV^{\onesite}$ of the corresponding bond-dimension $D$ MPS manifold, $\mc{M}_{D\text{-MPS}}$.
Since $\mc{M}_{D\text{-MPS}}$ has a curvature, it is not possible to perform generic non-infinitesimal rotations of $\ket{\Psi}$ in $\doubleV^{\onesite}$ without leaving the manifold $\mc{M}_{D\text{-MPS}}$.
To perform non-infinitesimal updates without leaving $\mc{M}_{D\text{-MPS}}$, the single-site DMRG and TDVP algorithms exploit that $\mc{M}_{D\text{-MPS}}$ has extensive flat
regions, contained in the \textit{local subspaces} $\doubleV[\ell]^{\onesite}$ of the tangent space $\doubleV^{\onesite}$~\cite{Haegeman2014}. 

Thus, we can successively update $\ket{\Psi}$ by sweeping through the subspaces $\doubleV[\ell]^{\onesite}$ and performing non-infinitesimal rotations in these spaces,
without leaving $\mc{M}_{D\text{-MPS}}$.
Each $\doubleV[\ell]^{\onesite}$ is the image of a local projector $\mc{P}^{\onesite}_{\ell}$, and $\doubleV^{\onesite}$ is the image of the tangent space projector
$\mc{P}^{\onesite} = \sum_{\ell=1}^{\scripteLL} \mc{P}^{\onesite}_{\ell} - \sum_{\ell=1}^{\scripteLL-1} \mc{P}^{\mbond}_{\ell}$, where the subtraction of the bond projectors
$\mc{P}^{\mbond}_{\ell}$ accounts for the fact that the $\doubleV[\ell]^{\onesite}$ spaces are not mutually orthogonal.

 By sweeping once from $\ell = 1$ to $\ell = \eLL$ or vice versa, we cover all possible directions in $\doubleV^{\onesite}$. 
For TDVP, $\ket{\Psi}$ should be rotated along every direction in $\doubleV^{\onesite}$ only once per sweep. 
Since the spaces $\doubleV[\ell]^{\onesite}$ are not mutually orthogonal, any rotations in $\doubleV[\ell]^{\onesite}$ in directions 
that are also present in still-to-be updated spaces $\doubleV[\ell']^{\onesite}$ have to be reversed again to avoid overcounting. 
For TDVP, this is taken care of by the backward time evolution steps when using the projector-splitting integrator of Ref.~\onlinecite{Haegeman2016}. 
 For DMRG, by contrast, considering certain directions multiple times during a sweep is not problematic. 

It should be noted that an update at some site $\ell$ of course changes $\ket{\Psi}$ and therefore its tangent space $\doubleV^{\onesite}$. 
Thus, subsequent updates, for instance, at site $\ell-1$, consider subspaces of an \textit{updated} tangent space. In DMRG, this is usually not an 
issue. In TDVP on the other hand, it leads to a Trotter error, which can be reduced by either using smaller time steps $\delta$ or by 
employing higher order Trotter schemes~\cite{Haegeman2016}. Both $\delta$ and the Trotter order control the number of sweeps necessary to time-evolve
from some initial time $t_0$ to a final time $t_1$. 

For both time evolution and ground-state search, it is ultimately relevant to consider changes of $\ket{\Psi}$ when
acted upon with the Hamiltonian $H$, $\ket{\delta \Psi} = (H-E) \ket{\Psi}$, where
$E$ is the energy expectation value of $\ket{\Psi}$. In single-site algorithms, $\ket{\delta \Psi}$ is projected to the tangent space,
i.e. $\ket{\delta \Psi^{\onesite}} = \mc{P}^{\onesite} \ket{\delta \Psi}$. These algorithms are therefore subject to a projection 
error $\ket{\delta \Psi^{\perp}} = (\doubleI - \mc{P}^{\onesite}) \ket{\delta \Psi}$. 

Both the standard 2s update and the CBE update aim to reduce the \textit{projection} error by incorporating new directions from the 2s space $\doubleV^{\twosite}$ which are not present in $\doubleV^{\onesite}$,
i.e.\ directions in $\doubleV^{2\perp}$, the image of $\mc{P}^{2\perp} = \sum_{\ell=1}^{\scripteLL-1} \mc{P}^{\ds\ds}_{\ell,\ell+1}$.
The corresponding 2s projection error (that would be ignored if these directions were neglected) is given by 
$\ket{\delta \Psi^{2\perp}} = \mc{P}^{2\perp} \ket{\delta \Psi} = \mc{P}^{2\perp} H \ket{\Psi}$.
In a similar spirit to the sweeping scheme used in the one-site update, $\doubleV^{2\perp}$ can be decomposed 
into local subspaces $\doubleV[\ell]^{2\perp}$, the images of the projectors $\mc{P}^{\ds\ds}_{\ell,\ell+1}$.
In 2s-DMRG, these are treated successively. 
While the 2s update considers the full space $\doubleV[\ell]^{2\perp}$, the main insight leading to CBE is that 2s accuracy can be achieved by 
considering only a small subspace of $\doubleV[\ell]^{2\perp}$.
This relevant subspace is identified by performing an (approximate) SVD on $\mc{P}^{\ds\ds}_{\ell,\ell+1} H \ket{\Psi}$, followed, 
in a right-to-left sweep, by an expansion of the isometry $A_{\ell}$ of $\ket{\Psi}$~\cite{Gleis2023a}.
This expansion reduces the contribution of $\mc{P}^{\ds\ds}_{\ell,\ell+1} \ket{\delta \Psi}$ to the 2s projection error. 

\vspace{-0.5cm}
\subsection{Importance of the $\mc{P}^{\ds\ds}_{\ell,\ell+1}$ projection}
\label{subsec:DDimportance}
\vspace{-0.2cm}

To be concrete, consider updating site $\ell+1$ during a right-to-left sweep. 
Before the update, the CBE algorithm expands the bond between sites $\ell$ and $\ell+1$,
selecting expansion vectors based on $\mc{P}^{\ds\ds}_{\ell,\ell+1} \ket{\delta \Psi}$.
In their comment, MO 
question this choice, and suggest to instead omit
the $\D$ projection at site $\ell+1$, i.e.\ to select expansion vectors based on $(\mc{P}^{\ds\ds}_{\ell,\ell+1} + \mc{P}^{\ds\ks}_{\ell,\ell+1}) \ket{\delta \Psi}$.
MO claim that the $\D$ projection at site $\ell+1$ discards useful information and can lead to problems, especially in TDVP. 
Below, we explain why this claim is incorrect. 
We also provide examples from TDVP that illustrate that the omission of the $\D$ projection at site $\ell+1$
can lead to avoidable projection errors.

To understand the difference between CBE and MO's suggestion, the following
projector identities are useful:
\begin{align}
\label{eq:PDD}
\raisebox{-0.57cm}{\includegraphics{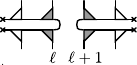}} & 
= \mc{P}^{\ds\ds}_{\ell,\ell+1}  = \mc{P}^{\twosite}_{\ell} (\doubleI - \mc{P}^{\onesite}) , 
\\
\label{eq:PDone}
\raisebox{-0.57cm}{\includegraphics{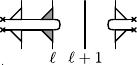}} &   
= \mc{P}^{\ds \ks}_{\ell, \ellplustwo}
= \mc{P}^{\twosite}_{\ell} (\doubleI - \mc{P}^{\onesite}_{\ell+1}) 
\\ \label{eq:PDoneTwo}
& \hspace{-2.2cm} = \raisebox{-0.57cm}{\includegraphics{Eq/PDDprojector}} + \raisebox{-0.57cm}{\includegraphics{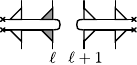}} = \mc{P}^{\ds\ds}_{\ell,\ell+1} + \mc{P}^{\ds\ks}_{\ell,\ell+1} . 
\vspace{-0.1cm}
\end{align}
They follow from  Eqs.~(33), (37), (53), (54)  of Ref.~\cite{Gleis2022a}; the diagrams illustrate them for a chain of length $\eLL = 4$.

CBE ``targets'' states (i.e.\ includes them during 
bond expansion) from the image of the projector 
$\mc{P}^{\ds\ds}_{\ell,\ellplusone}$ of Eq.~\eqref{eq:PDD}.
The right side of \eqref{eq:PDD} shows that this
reduces the 1s tangent space projection error $\ket{\delta \Psi^{\perp}} = (\doubleI - \mc{P}^{\onesite}) \ket{\delta \Psi}$
by reducing its 2s part by a local contribution lying in the image
of $\mc{P}^\twosite_\ell$. 
Therefore, CBE targets only directions \text{not} present in the 
1s tangent space $\doubleV^{\onesite}$, 
i.e.\ directions that cannot be accessed during a full 
1s sweep.

MO's suggestion, by contrast, targets states in the image of 
the projector given by Eq.~\eqref{eq:PDone}, which differs 
from \eqref{eq:PDone} by having $\doubleI$ instead of a $\D$ projection at site $\ellplusone$. The right side of \eqref{eq:PDone} shows that this reduces the \textit{local}  projection error
$\ket{\delta \Psi^{\perp}_{\ell+1}} = (\doubleI - \mc{P}_{\ell+1}^{\onesite}) \ket{\delta \Psi}$ via a local 2s contribution
in the image of $\mc{P}^\twosite_\ell$. 
Moreover, expressing  MO's projector in the form \eqref{eq:PDoneTwo}, we see that it targets not only the CBE target states in 
the image of $\mc{P}^{\ds \ds}_{\ell, \ellplusone}$,
but also states in the 
image of $\mc{P}^{\ds \ks}_{\ell, \ellplusone}$. Now, 
the image of $\mc{P}^{\ds \ks}_{\ell, \ellplusone}$ is a subspace of the local space $\doubleV[\ell]^{\onesite} \subset \doubleV^{\onesite}$, 
which will be dealt with anyway during the 1s update at site $\ell$. In other words, MO's suggestion differs from CBE by also targeting directions already present in $\doubleV^{\onesite}$. This is not efficient: if the number of expansion vectors is limited (as is the case in practice), it is important to focus on directions that cannot be accessed during a full sweep, i.e.\ directions which are not present in $\doubleV^{\onesite}$. 
CBE therefore purposefully projects out directions present in $\doubleV^{\onesite}$.

For DMRG, the inclusion of states from 
$\mc{P}^{\ds\ks}_{\ell,\ell+1}H \ket{\Psi}$
during bond expansion,
though not efficient, will not lead to serious problems, since DMRG minimizes the one-site variance \cite{Hubig2018} and thus the norm of $\mc{P}_{\ell,\ell+1}^{\ds\ks} \ket{\delta \Psi}$. Close to convergence, the latter
becomes very small, so that  $(\mc{P}_{\ell,\ell+1}^{\ds\ds} + \mc{P}_{\ell,\ell+1}^{\ds\ks})\ket{\delta \Psi} \simeq \mc{P}_{\ell,\ell+1}^{\ds\ds}\ket{\delta \Psi}$, i.e.\ the expansion 
strategies of CBE and MO become equivalent. 

For TDVP, by contrast, targeting states from 
the image of $\mc{P}_{\ell,\ell+1}^{\ds\ks} + \mc{P}_{\ell,\ell+1}^{\ds\ds}$ instead of $\mc{P}_{\ell,\ell+1}^{\ds\ds}$
can cause not only
(i) avoidable computational costs, but also, in the worst case (ii) avoidable projection errors. In the following, we elaborate (i) and (ii) through simple but explicit examples.

(i) Avoidable computational costs are generated if, 
during a TDVP time step update of site $\ellplusone$,
bond expansion includes directions already present in $\doubleV[\ell]^{\onesite} \subset \doubleV^{\onesite}$. The update of site $\ellplusone$ then includes rotations of $\ket{\Psi}$ into these directions. However,
since these directions will also be considered when subsequently updating site $\ell$, TDVP's backward time step will undo rotations along them before moving on to site $\ell$. Thus,
the inclusion of such direction in bond expansion is computationally inefficient. 

As an example, consider  $H_{1} = \ket{10}\bra{00} + \ket{00}\bra{10}$ as Hamiltonian, with local space $\doublev[\,] = \mr{span}\{\ket{0},\ket{1}\}$. Take $\ket{00}$ as initial state and use a right-to-left sweep to perform a single TDVP time step.
The relevant projectors are 
\begin{align}
\mc{P}^{\onesite}_1 &= \doubleI \otimes \ket{0}\bra{0} \, , &\mc{P}^{\onesite}_2 &= \ket{0}\bra{0} \otimes \doubleI \nonumber \\
\mc{P}^{\mbond}_1 &= \ket{0}\bra{0} \otimes \ket{0}\bra{0} \, , &\mc{P}^{\ds\ds}_{1,2} &= \ket{1}\bra{1} \otimes \ket{1}\bra{1} \nonumber\\
\mc{P}^{\ds\ks}_{1,2} &= \ket{1}\bra{1} \otimes \ket{0}\bra{0} \, , &\mc{P}^{\ks\ds}_{1,2} &= \ket{0}\bra{0} \otimes \ket{1}\bra{1} \nonumber 
\, .
\end{align}

Since $H_1 \ket{00} = \ket{10}$ and therefore $\mc{P}^{\ds\ds}_{1,2} H_1 \ket{00} = 0$, CBE--TDVP does not 
perform bond expansion in this case. The update at site $2$, which is based on $\mc{P}^{\onesite}_2 H_1 \mc{P}^{\onesite}_2 = 0$, does nothing
and yields $\ket{00}$. Likewise, the subsequent bond update based on $\mc{P}^{\mbond}_1 H_1 \mc{P}^{\mbond}_1 = 0$ does nothing and yields $\ket{00}$. 
Finally, the update on site 1, which is based on $\mc{P}^{\onesite}_1 H_1 \mc{P}^{\onesite}_1 = H_1$,
gives the correct result $\ket{00} - \mi \delta \ket{10} + \mc{O}(\delta^2)$, where $\delta$ is the time step.
We have omitted $\mc{O}(\delta^2)$ contributions since these are subject to the Trotter error,
which is not controlled by bond expansion.
To summarize, CBE--TDVP generates the following sequence of states: 
\begin{align*}
\ket{00} \overset{\text{site 2}}{\to} \ket{00} \overset{\text{bond}}{\to} \ket{00} \overset{\text{site 1}}{\to} \ket{00} - \mi \delta  \ket{10} \, .
\end{align*}

If we follow MO's suggestion of omitting the $\D$ projection, the bond is expanded, since $(\mc{P}^{\ds\ks}_{1,2}  + \mc{P}^{\ds\ds}_{1,2}) H_1 \ket{00} = \ket{10}$.
Thus, the isometry $A_1$ is expanded by $\ket{1}$, and $[H_1]^{\onesite,\mr{ex}}_{2} = H_1$. We therefore generate the sequence
\begin{align*}
\ket{00} \overset{\text{site 2}}{\to} \ket{00} - \mi \delta \ket{10} \overset{\text{bond}}{\to} \ket{00} \overset{\text{site 1}}{\to} \ket{00} - \mi \delta \ket{10} \, .
\end{align*}
The final state is again correct, but we have unnecessarily evolved the state back and forth in time, while additionally also using a larger bond dimension. 

(ii) In almost all practical cases, we have to select a small number of expansion vectors from a very large set of possible candidates.
Then, MO's suggestion of omitting the $\D$ projection at site $\ell+1$ may cause  
some vectors from $\mc{P}_{\ell,\ell+1}^{\ds\ks}\ket{\delta \Psi}$ 
to be included at the cost of omitting some vectors from 
$\mc{P}_{\ell,\ell+1}^{\ds\ds}\ket{\delta \Psi}$. 
That is not only inefficient but can also lead to projection errors
that could have been avoided otherwise.

As an example, consider $H_2 = H_1  + \omega_0 [ \ket{22}\bra{00} + \ket{00}\bra{22}]$, with $\omega_0 < 1$.
The local space is now $\doublev[\,] = \mr{span}\{\ket{0},\ket{1},\ket{2}\}$.
Again, we perform a right-to-left sweep starting from $\ket{00}$. Suppose that during bond expansion, we limit the number of expansion 
vectors to 1, i.e. we only select a single expansion vector. This serves as a toy model for realistic situations, where
we constantly have to choose a small set of expansion vectors from a very large set.
In the present case, CBE-TDVP chooses $\ket{2}$ to expand the isometry $A_1$, which is the only available choice in $\DD$,
and generates the following sequence:
\begin{align*}
\ket{00} &\overset{\text{site 2}}{\to} \ket{00} - \mi\delta \omega_0 \ket{22}  \overset{\text{bond}}{\to} \ket{00} \\
&\overset{\text{site 1}}{\to} \ket{00} - \mi\delta [\ket{10} + \omega_0\ket{22}] \, .
\end{align*}
This is the correct result to order $\delta$, i.e.\ CBE--TDVP avoids the projection error by appropriate bond expansion.
If we follow MO's suggestion and omit the $\D$ projection at site 2, we choose $\ket{1}$ to expand $A_1$, since $\omega_0 < 1$. 
We then generate the sequence
\begin{align*}
\ket{00} &\overset{\text{site 2}}{\to} \ket{00} - \mi\delta \ket{10}  \overset{\text{bond}}{\to} \ket{00} \overset{\text{site 1}}{\to} \ket{00} - \mi\delta  \ket{10} \, .
\end{align*}
This result differs from the correct result at order $\delta$, since we missed the $\ket{22}$ contribution.

\end{document}